\documentclass[12pt]{article}
\usepackage{color}
\usepackage{amssymb,amsmath,comment}
\usepackage{graphicx}
\usepackage{braket}
\usepackage{epsfig}
\usepackage{here}
\usepackage{bm}
\graphicspath{{./Figures/}}
\newcommand{\ba}{\begin{align}}
\newcommand{\ea}{\end{align}}

\def\nn{\nonumber}

\def\bea{\begin{eqnarray}}
\def\eea{\end{eqnarray}}
 \makeatletter
\def\alt{\mathrel{\mathpalette\gl@align<}}
\def\agt{\mathrel{\mathpalette\gl@align>}}
\def\gl@align#1#2{\lower.6ex\vbox{\baselineskip\z@skip\lineskip\z@
\ialign{$\m@th#1\hfil##\hfil$\crcr#2\crcr\sim\crcr}}} \makeatother
\renewcommand{\thefootnote}{\fnsymbol{footnote}}
\begin{document}
\begin{flushright}
\end{flushright}
\vspace*{1.0cm}

\begin{center}
\baselineskip 20pt 
{\Large\bf 
Revisiting Quantum Stabilization of
\\
the Radion in Randall-Sundrum Model
}
\vspace{1cm}

{\large 
Naoyuki Haba$^a$
\ and \ Toshifumi Yamada$^b$
} \vspace{.5cm}

{\baselineskip 20pt \it
$^a$ Institute of Science and Engineering, Shimane University, Matsue 690-8504, Japan\\
$^b$ Department of Physics, Kyoto University, Kyoto, Kyoto 606-8502, Japan
}

\vspace{.5cm}

\vspace{1.5cm} {\bf Abstract} \end{center}

We study the stabilization of the radion in Randall-Sundrum-1 model
 by the Casimir energy of a bulk gauge field.
The Casimir energy is proportional to a divergent, infinite summation over the zeros of a Wronskian of Bessel functions that implicitly depends on the radion vacuum expectation value,
 and its regularization and renormalization is the central issue.
We carry out the correct regularization and renormalization by noting that
 analytic continuation must be performed only on functions that are independent of the radion vacuum expectation value.
Thereby we find that the 1-loop effective potential of the radion generated by the Casimir energy
 can be renormalized with the boundary tensions,
 and we correctly obtain the renormalized effective potential.
It is shown that a bulk gauge field satisfying Dirichlet condition at the positive (UV) boundary and Dirichlet condition at the negative (IR) boundary gives rise to an appropriate radion potential that stabilizes the radion vacuum expectation value in a way that a large hierarchy of the warp factor is generated naturally.

\thispagestyle{empty}

\newpage
\renewcommand{\thefootnote}{\arabic{footnote}}
\setcounter{footnote}{0}
\baselineskip 18pt
\section{Introduction}

The Randall-Sundrum-1 (RS-1) model~\cite{Randall:1999ee} offers an intriguing solution to the big hierarchy problem between 10~TeV scale, 
 where the cutoff of the standard model is expected to exist, and the scale just below the Planck mass.
One issue in RS-1 model is the stabilization of the radion.
The radion is the scalar degree of freedom of the spacetime metric in RS-1 model,
 whose vacuum expectation value (VEV) regulates the distance between the two boundaries
\footnote{
Throughout the paper, we adopt the interval picture~\cite{Lalak:2001fd,Carena:2005gq}, 
 where the 5th dimension is an interval $r\leq y\leq0$
 and two boundaries are located at $y=0,r$.
}
 and thus determines the amount of redshift at the IR boundary.
In the original RS-1 model, the radion is massless.
Hence, we need an external mechanism to stabilize the radion VEV, 
 and in order for RS-1 model to be a solution to the big hierarchy problem,
 the stabilization must be achieved without fine-tuning of the relevant renormalization constant.

In this paper, we investigate a mechanism for the radion stabilization
 which utilizes the Casimir energy of a bulk field.
This is an alternative to the well-studied Goldberger-Wise mechanism~\cite{Goldberger:1999uk},
 which exploits a bulk scalar field and boundary-localized classical potentials, in contrast to quantum (Casimir) effects that we use.
This paper concentrates on the case with a bulk gauge field~\cite{Davoudiasl:1999tf,Pomarol:1999ad},
 because its bulk mass is forbidden by the gauge principle and hence one can make more restrictive predictions than the case with a bulk scalar or fermion.

Formerly, evaluation of the Casimir energy of a bulk field in 5D RS-1 model with 4D flat branes
 has been carried out in Refs.~\cite{Goldberger:2000dv,Toms:2000vm,Flachi:2001pq,Flachi:2001ke,Flachi:2001bj,Garriga:2000jb,Garriga:2002vf,Brevik:2000vt,Saharian:2002bw,Knapman:2003ey}.
In those works, analytic continuation (in the form of zeta function regularization or dimensional regularization) has been used to regularize and renormalize
 the divergent Casimir energy.
In the present paper, we re-calculate the Casimir energy, and the 1-loop effective potential for the radion generated by it, 
 by respecting the following principle:
\bea
&&Functions \ to \ be \ analytically \ continued \ must \ not \ depend \ on \ the \ radion \ VEV.
\label{principle}
\eea
\noindent
We argue that this principle is mandatory for the correct regularization and renormalization of the radion effective potential.
To see this, suppose one has a function, $F(s,\,a_{\rm rad})$, that depends both on a parameter $s$ and the radion-VEV-dependent warp factor $a_{\rm rad}$ and that is defined only for Re$(s)<s_0$.
One wants to know a regularized-and-renormalized value of $F$ at $s$ with Re$(s)\geq s_0$ in order to evaluate the radion effective potential.
We argue that one must \textit{not} perform the analytical continuation of $F(s,\,a_{\rm rad})$
 from Re$(s)<s_0$ to Re$(s)\geq s_0$,
 because physically, analytical continuation is a rule to relate a divergent quantity to a finite quantity whose uniqueness is guaranteed by the identity theorem.
Since $F$ with different values of the radion VEV, $F(s,\,a'_{\rm rad})$
 and $F(s,\,a''_{\rm rad})$, are different functions of $s$,
 the identity theorem does not hold, and the relation between a divergent quantity and a finite quantity is not unique and can depend on the radion VEV.
This is an incorrect regularization and renormalization, because the difference between the divergent and finite quantities may depend on $a_{\rm rad}$ in a way that it cannot be renormalized with the boundary tensions, but it is wrongly subtracted in the analytically-continued $F$.
To avoid the above error, analytic continuation must be performed only on $a_{\rm rad}$-independent functions.

To satisfy the principle~(\ref{principle}), we express an infinite summation
 over the zeros of a Wronskian of Bessel functions, in terms of an elaborate contour integral in which the functions that need to be analytically continued are independent of the radion VEV.
This contour integral is a generalization of Abel-Plana formula,
 which may be akin to the one in Ref.~\cite{Saharian:2007ph}.
Using the elaborate contour integral, and abiding by (\ref{principle}),
 we perform the regularization and renormalization of the 1-loop effective potential of the radion.
We show that this potential can be renormalized with the boundary tensions, and obtain the renormalized 1-loop effective potential of the radion.
Finally, we examine if the renormalized potential realizes radion stabilization without fine-tuning of the renormalization constant.
 \\

This paper is organized as follows:

In Section~2, we present the action for the gravity part of RS-1 model in the interval picture~\cite{Lalak:2001fd,Carena:2005gq}, 
 and study the equations of motion following from it.

In Section~3, we re-derive the spacetime metric containing the radion fluctuation, using the interval picture.
This section is slightly besides the main topic of the present paper,
 but we include it considering the importance of the accurate definition of the radion.
In the Gaussian normal coordinate with respect to one boundary, we solve the linearized bulk Einstein equation and equation of motion at that boundary, to derive the radion solution. We then argue that the other boundary is situated in a way that the radion solution satisfies the equation of motion there, and thereby determine the configuration of this boundary.
The radion solution has originally been derived in Ref.~\cite{Charmousis:1999rg}.
We reformulate that work in the interval picture and also aim at a more systematic derivation of the radion solution.
The radion in the interval picture has been studied in Ref.~\cite{Carena:2005gq}.
Our approach is different in that we start from the Gaussian normal coordinate with respect to one boundary,
 and then determine the configuration of the other boundary by requiring the existence of the radion.

In Subsection~4.1, we derive the expression for the Casimir energy of a bulk gauge field.

Subsections~4.2 and 4.3 are our new finding and the heart of this paper.
We evaluate the 1-loop effective potential of the radion generated by the Casimir energy, 
 by performing the regularization and renormalization of the infinite summation over the zeros of a Wronskian of Bessel functions respecting the principle~(\ref{principle}).
We present the numerical calculation of the correctly regularized-and-renormalized 1-loop effective potential of the radion
 and discuss radion stabilization therewith.

Section~5 is devoted to the conclusion.
\\

Throughout the paper, capital Roman letters $M,N,A,B,...$ indicate 5D spacetime coordinates,
 small Greek letters $\mu,\nu,\alpha,\beta,...$ 4D spacetime coordinates, and `5' explicitly points at the 5th dimension coordinate.
\\


\section{Randall-Sundrum-1 Model in the Interval Picture}

Consider a 5D spacetime given as an interval along the 5th dimension $y$, $0\leq y\leq r$, that has boundaries at $y=0$ and $y=r$.
The bulk has a negative bulk cosmological constant, $\Lambda=-\frac{1}{2}d(d+1)k^2M^3$.
The boundary at $y=0$ (called `positive boundary') has a fine-tuned positive tension, $V_{\rm UV}=d\, k\,M^3$,
 and the boundary at $y=r$ (called `negative boundary') has a fine-tuned negative tension, $V_{\rm IR}=-d\, k\,M^3$.
Here, $d$ denotes the dimension of space, with $d=3$ corresponding to our case.

The action for the gravity part is given by
\footnote{
In this paper, we neglect boundary-localized curvatures.
}
\begin{align}
S_{\rm grav}&=M^d\int_0^r{\rm d}y\int{\rm d}^{d+1}x\,\sqrt{-G}
\left\{-R+d(d+1)k^2\right\}
\nn\\
&+M^d\left.\int{\rm d}^{d+1}x\,\sqrt{-g}\left(2K-2d\ k\right)\right\vert_{y=0}
+M^d\left.\int{\rm d}^{d+1}x\,\sqrt{-g}\left(-2K+2d\ k\right)\right\vert_{y=r}
\label{action}
\end{align}
 where $G_{MN}$ is the metric of 5D spacetime, $g_{\mu\nu}$ is the induced metric on $y=$(constant) surfaces, $M$ is the 5D Planck mass, and $R$ is the scalar curvature.
$K=g^{\mu\nu}K_{\mu\nu}$ is the trace of the extrinsic curvature $K_{\mu\nu}$ on a $y=$(constant) surface.
The terms with $K$ are Gibbons-Hawking terms~\cite{York:1972sj,Gibbons:1976ue} which ensure that the correct Einstein's equation in the bulk
  is obtained from variational principle in which the metric at the boundaries is fixed, i.e. $\delta G_{MN}\vert_{y=0,r}=0$, but its derivative with $y$ can be non-zero, i.e. $\partial_y\delta G_{MN}\vert_{y=0,r}\neq0$.

The equation of motion of the metric is derived from variational principle in which
 we take $\delta G_{MN}\vert_{y=0,r}\neq0$ and thus the Gibbons-Hawking terms no longer cancel the boundary terms.
This leads to non-vanishing equations of motion at the boundaries $y=0,r$.
From variational principle, we get
\begin{align} 
0=\delta S_{\rm grav}&=
M^d\int_0^r{\rm d}y\int{\rm d}^{d+1}x\,\sqrt{-G}\ \delta G_{MN}\left\{R^{MN}-\frac{1}{2}G^{MN}R+G^{MN}\frac{1}{2}d(d+1)k^2\right\}
\label{variational1}\\
&+M^d\int{\rm d}^{d+1}x\,\sqrt{-g}\left\{ \, n_M\left(G^{AB}\delta\Gamma_{AB}^M-G^{MA}\delta\Gamma_{AB}^B\right)\right.
\nn\\
& \ \ \ \ \ \ \ \ \ \ \ \ \ \ \ \ \ \ \ \ \ \  \left. \left.+\delta g_{\mu\nu}\left(-2K^{\mu\nu}+g^{\mu\nu}K-g^{\mu\nu}d\ k\right)+2g^{\mu\nu}\delta K_{\mu\nu}
\,\right\}\right\vert_{y=0}
\label{variational2}\\
&+M^d\int{\rm d}^{d+1}x\,\sqrt{-g}\left\{\, -n_M\left(G^{AB}\delta\Gamma_{AB}^M-g^{MA}\delta\Gamma_{AB}^B\right)\right.
\nn\\
& \ \ \ \ \ \ \ \ \ \ \ \ \ \ \ \ \ \ \ \ \ \  \left. \left.+\delta g_{\mu\nu}\left(2K^{\mu\nu}-g^{\mu\nu}K+g^{\mu\nu}d\ k\right)-2g^{\mu\nu}\delta K_{\mu\nu}
\,\right\}\right\vert_{y=r}
\label{variational3}
\end{align}
 where $n_M$ is the unit vector transverse to a $y=$(constant) surface along $+y$ direction, satisfying $n_M n^M=-1$.
 $\delta\Gamma_{AB}^C$ is the variation of a Christoffel symbol resulting from $\delta g_{MN}$, and
 the first two terms of Eqs.~(\ref{variational2}),(\ref{variational3}) come from the total derivative term in the bulk $\sqrt{-g}g^{MN}\delta R_{MN}$.
Eq.~(\ref{variational1}) yields usual Einstein's equation in the bulk.
Eqs.~(\ref{variational2}),(\ref{variational3}) give the equations of motion at the boundaries that play the role of boundary conditions,
 replacing junction conditions~\cite{Israel:1966rt} in the orbifold picture.
\\

\section{Radion Solution}

The radion field is a fluctuation of spacetime metric off RS-1 spacetime,
 and is therefore a solution to the linearized bulk Einstein equation and boundary equations of motion.
To find out the radion solution, we take two steps:

\begin{enumerate}

\item We take a Gaussian normal coordinate with respect to the positive boundary.
In this coordinate, we solve the linearized bulk Einstein equation and equation motion at the positive boundary, to derive the radion solution.
However, we temporarily ignore the negative boundary.

\item Next, we take the negative boundary into consideration.
We do not consider that the negative boundary is located somewhere a priori.
Rather, we \textit{situate} the negative boundary in a way that the radion solution of Step~1 obeys the equation of motion at the negative boundary. We conjecture that a more fundamental theory that regulates dynamics of boundaries will justify the above procedure.
In this way, we determine the configuration of the negative boundary.
The radion solution of Step~1 is automatically promoted to a solution to all the equations of motion.

\end{enumerate}

\subsection{Step 1}

Take a Gaussian normal coordinate with respect to the positive boundary, $(\tilde{x}^\mu,\tilde{y})$.
The positive boundary is at $\tilde{y}=0$.
We quantify the metric fluctuations by $h_{\mu\nu}(\tilde{x},\tilde{y})$, as
\begin{align}
{\rm d}s^2=\left(e^{-2A(\tilde{y})}\,\eta_{\mu\nu}+h_{\mu\nu}(\tilde{x},\tilde{y})\right){\rm d}\tilde{x}^\mu{\rm d}\tilde{x}^\nu-{\rm d}\tilde{y}^2,
\label{gn0}
\end{align}
 where $A$ is a function of only $\tilde{y}$. 
\footnote{
We raise and lower the indices of $h_{\mu\nu}$ with the Minkowski metric $\eta_{\mu\nu}$.
We also define $h=h_{\mu\nu}\eta^{\mu\nu}$.
}

We comment on the gauge fixing of $h_{\mu\nu}(\tilde{x},\tilde{y})$.
The only coordinate transformations that maintain the Gaussian normal coordinate (i.e. keep $G_{\mu 5}=0$ and $G_{55}=-1$)
 and that do not change the boundary position, are $\tilde{x}^\mu\to \tilde{x}^\mu+\epsilon^\mu(\tilde{x})$ with $\epsilon^\mu(\tilde{x})$ being a function of only $\tilde{x}^\mu$.
So, $h_{\mu\nu}(\tilde{x},\tilde{y})$ can only transform as
\bea
h_{\mu\nu}(\tilde{x},\tilde{y}) \ \to \ h_{\mu\nu}(\tilde{x},\tilde{y})-e^{-2A(\tilde{y})}\left( \partial_\mu\epsilon_\nu(\tilde{x})+\partial_\nu\epsilon_\mu(\tilde{x})\right).
\label{trans}
\eea
Since $h_{\mu\nu}(\tilde{x},\tilde{y})$ is a general function of $\tilde{y}$, 
 we cannot impose any gauge fixing condition using Eq.~(\ref{trans})
 unless we solve the equations of motion and specifies its $\tilde{y}$-dependence.

To the zeroth and first orders of $h_{\mu\nu}$, the bulk Einstein equation gives (we write $A'=\frac{{\rm d}A}{{\rm d}\tilde{y}}$)
\begin{align}
R_{\mu\nu}=&(d+1)k^2G_{\mu\nu}\Rightarrow
\ \ \ -A''+(d+1)A'^2=(d+1)k^2,
\label{eom1}\\
& \ \ \ \ \ \ \ \ \ \ \ \ \ \ \ \ \ \ \ \ \ \ \ \   \frac{1}{2}e^{2A}\left(\partial_\mu\partial_\alpha h^{\alpha}_\nu+\partial_\nu\partial_\alpha h^{\alpha}_\mu-\square h_{\mu\nu}-\partial_\mu\partial_\nu h\right)
\nn\\
&\ \ \ \ \ \ \ \ \ \ \ \ \ \ \ \ \  +\frac{1}{2}\frac{\partial^2 h_{\mu\nu}}{\partial \tilde{y}^2}+\frac{3-d}{2}\frac{\partial h_{\mu\nu}}{\partial \tilde{y}}+2A'^2h_{\mu\nu}
-\frac{1}{2}A'\eta_{\mu\nu}\frac{\partial h}{\partial \tilde{y}}-A'^2\eta_{\mu\nu}h=(d+1)k^2h_{\mu\nu}.
\label{eom2}\\
R_{\mu5}=&0\Rightarrow \ \ \ \ \ \left(\frac{1}{2}\frac{\partial}{\partial \tilde{y}}+A'\right)\left(\partial_\alpha h^{\alpha}_\mu-\partial_\mu h\right)=0.
\label{eom3}\\
R_{55}=&(d+1)k^2G_{55}\Rightarrow \ \ \ A''-A'=-k^2,
\label{eom4}\\
& \ \ \ \ \ \ \ \ \ \ \ \ \ \ \ \ \ \ \ \ \ \ \ \frac{1}{2}\frac{\partial^2h}{\partial \tilde{y}^2}+A'\frac{\partial h}{\partial \tilde{y}}+A''h=0.
\label{eom5}
\end{align}
The equation of motion at the positive boundary following from Eq.~(\ref{variational2})
 is vastly simplified in the present coordinate and becomes
\begin{align}
&\left.\frac{1}{2}\frac{\partial g_{\mu\nu}}{\partial \tilde{y}}-\frac{1}{2}g_{\mu\nu}g^{\alpha\beta}\frac{\partial g_{\alpha\beta}}{\partial \tilde{y}}-g_{\mu\nu}\ d\ k\right\vert_{\tilde{y}=0}=0,
\label{beom}
\end{align}
 which gives, to the zeroth and first orders of $h_{\mu\nu}$,
\begin{align}
&A'(\tilde{y}=0)-k=0,
\label{beom1}\\
&\left.\frac{1}{2}\frac{\partial h_{\mu\nu}}{\partial \tilde{y}}+\left\{(1+d)A'-d\ k\right\}h_{\mu\nu}
-\frac{1}{2}\eta_{\mu\nu}\frac{\partial h}{\partial \tilde{y}}-A' \, \eta_{\mu\nu}h\right\vert_{\tilde{y}=0}=0.
\label{beom2}
\end{align}

We solve the equations of motion~(\ref{eom1})-(\ref{beom2}). One easily confirms $A(\tilde{y})=k\tilde{y}$.
To solve for $h_{\mu\nu}$, we go to the 4D momentum space and write the 4D momentum of $h_{\mu\nu}$ as $p^\mu$.
The treatment of the equations of motion is different for $p^2\neq0$ (massive) case and $p^2=0$ (massless) case, which we discuss separately below:
\begin{itemize}

\item The $p^2\neq0$ case:

We decompose $h_{\mu\nu}$ as
\begin{align}
h_{\mu\nu}(p,\tilde{y})=t_{\mu\nu}(p,\tilde{y})+p_\mu\,V_\nu(p,\tilde{y})+p_\nu\,V_\mu(p,\tilde{y})&+p_\mu p_\nu \, S_1(p,\tilde{y})+\eta_{\mu\nu}\,S_2(p,\tilde{y})
\label{para1}
\\ &{\rm with} \ \ t^{\mu}_\mu=0, \ \ p^\mu t_{\mu\nu}=0, \ \ p^\mu V_\mu=0,
\nn
\end{align}
 where $t_{\mu\nu}$ is a transverse-traceless tensor, $V_\mu$ is a divergence-free vector, and $S_1,S_2$ are scalars.
Plugging Eq.~(\ref{para1}) into bulk equations~(\ref{eom3}),(\ref{eom5}) and the trace of boundary equation~(\ref{beom2}), we find
 the following $\tilde{y}$-dependence of the components:
\bea
&&p^\mu h_{\mu\nu}\propto e^{-2k\tilde{y}} \, \Rightarrow \, p^2\,V_\nu+p^2 p_\nu \, S_1+p_\nu \, S_2 \propto e^{-2k\tilde{y}},
\nn\\
&&h\propto e^{-2k\tilde{y}} \, \Rightarrow \, p^2\,S_1+(d+1)S_2 \propto e^{-2k\tilde{y}},
\nn
\eea
 from which we conclude $S_1,\,S_2,\,V_\nu\propto e^{-2k\tilde{y}}$.
Now that we have $S_1,\,V_\nu \propto e^{-2k\tilde{y}}$, we can utilize a coordinate transformation Eq.~(\ref{trans})
 to take the gauge with $S_1=0$ and $V_\nu=0$.
Inserting Eq.~(\ref{para1}) into bulk and boundary equations~(\ref{eom2}),(\ref{beom2}) with $S_1=0$ and $V_\nu=0$, and using $S_2\propto e^{-2k\tilde{y}}$,
 we arrive at
\begin{align}
&-\frac{1}{2}e^{2k\tilde{y}}\left\{-p^2\, t_{\mu\nu}+(1-d)p_\mu p_\nu\, S_2-p^2\eta_{\mu\nu} \, S_2\right\}+
\frac{1}{2}\frac{\partial^2 t_{\mu\nu}}{\partial \tilde{y}^2}+\frac{3-d}{2}k\frac{\partial t_{\mu\nu}}{\partial \tilde{y}}+(1-d)k^2\,t_{\mu\nu}=0,
\nn\\\label{massiveeom}
\\
&\left.\frac{1}{2}\frac{\partial t_{\mu\nu}}{\partial \tilde{y}}+2k\,t_{\mu\nu}\right\vert_{\tilde{y}=0}=0.
\end{align}
The solution to the above equations
 is $S_2=0$ and 
 $t_{\mu\nu} \propto e^{\frac{3-d}{2}k\tilde{y}}Y_{\frac{d-1}{2}}(\frac{p}{k})J_{\frac{d+1}{2}}(\frac{p}{ke^{-k\tilde{y}}})-e^{\frac{3-d}{2}k\tilde{y}}J_{\frac{d-1}{2}}(\frac{p}{k})Y_{\frac{d+1}{2}}(\frac{p}{ke^{-k\tilde{y}}})$,
 which manifests that there is no massive scalar and there exists a massive spin-2 field.

\item The $p^2=0$ case:

Because $p^2=0$, \ $t_{\mu\nu}$ and $V_\mu$ of Eq.~(\ref{para1}) now contain a longitudinal component proportional to $p^\mu$, 
 which contaminates other components and so must be isolated to make the parametrization well-defined.
To unambiguously isolate such components, we introduce a constant vector, $C^\mu$, satisfying $p_\mu C^\mu\neq0$.
We decompose $t_{\mu\nu}$ and $V_\mu$ as $t_{\mu\nu}=\tau_{\mu\nu}+p_\mu \alpha_\nu+p_\nu \alpha_\mu+p_\mu p_\nu \phi$,
 $V_\mu=U_\mu+W p_\mu$ such that $C^\mu \tau_{\mu\nu}=0$, \, $C^\mu \alpha_\mu=0$, \, $C^\mu U_\mu=0$.
The decomposition for $h_{\mu\nu}$ then becomes
\begin{align}
h_{\mu\nu}(p,\tilde{y})=&\tau_{\mu\nu}(p,\tilde{y})+p_\mu\left\{U_\nu(p,\tilde{y})+\alpha_\nu(p,\tilde{y})\right\}
\nn\\
&+p_\nu\left\{U_\mu(p,\tilde{y})+\alpha_\mu(p,\tilde{y})\right\}+p_\mu p_\nu\left\{S_1(p,\tilde{y})+\phi(p,\tilde{y})+W(p,\tilde{y})\right\}+\eta_{\mu\nu}\,S_2(p,\tilde{y})
\label{para2}
\\ & \ \ \ \ \ \ {\rm with} \ \ \tau^{\mu}_\mu=0, \ \ p^\mu \tau_{\mu\nu}=C^\mu \tau_{\mu\nu}=0, \ \ p^\mu \alpha_\mu=C^\mu \alpha_\mu=0,
 \ \ p^\mu U_\mu=C^\mu U_\mu=0.
\nn
\end{align}
Plugging Eq.~(\ref{para2}) into bulk equations~(\ref{eom3}),(\ref{eom5}) and boundary equation~(\ref{beom2}),
 we get 
\bea
p^\mu h_{\mu\nu}\propto e^{-2k\tilde{y}} \, \Rightarrow \, p_\nu \, S_2 \propto e^{-2k\tilde{y}}, \ \ \ \ \ \ \ \
h\propto e^{-2k\tilde{y}} \, \Rightarrow \,(d+1)S_2 \propto e^{-2k\tilde{y}},
\nn
\eea
 which only give $S_2\propto e^{-2k\tilde{y}}$ and do not provide any information on other components.
Thus, no gauge fixing can be performed at this stage.
Inserting Eq.~(\ref{para2}) into bulk and boundary equations~(\ref{eom2}),(\ref{eom5}) without gauge fixing, we find
\begin{align}
&-\frac{1}{2}e^{2k\tilde{y}}(1-d)p_\mu p_\nu\, S_2+\frac{1}{2}\left[\frac{\partial^2}{\partial \tilde{y}^2}+\frac{3-d}{2}k\frac{\partial}{\partial \tilde{y}}+(1-d)k^2\right]
\nn\\
&\ \ \ \ \ \ \ \ \ \ \ \ \ \ \ \ \ \ \ \ \ \ \times\left\{\tau_{\mu\nu}+p_\mu(U_\nu+\alpha_\nu)+p_\nu(U_\mu+\alpha_\mu)+p_\mu p_\nu(S_1+\phi+W)\right\}=0,
\label{masslesseom}
\\
&\left.\left(\frac{1}{2}\frac{\partial}{\partial \tilde{y}}+2k\right)\left\{\tau_{\mu\nu}+p_\mu(U_\nu+\alpha_\nu)+p_\nu(U_\mu+\alpha_\mu)+p_\mu p_\nu(S_1+\phi+W)\right\}
\right\vert_{\tilde{y}=0}=0.
\end{align}
The solution to the above equations is $\tau_{\mu\nu} \propto e^{-2k\tilde{y}}$, $U_\nu+\alpha_\nu\propto e^{-2k\tilde{y}}$,
 and two independent solutions for $S_1+\phi+W$ given by
\bea
S_1+\phi+W \,\propto\, e^{-2k\tilde{y}} \ \ \ \ \ \ {\rm (first \ solution)}
\label{sol1}
\eea
and
\bea
S_1+\phi+W = \frac{1}{k^2}\left(\frac{1}{2}-\frac{1}{d+1}e^{(d-1)k\tilde{y}}\right)f(p)\ \ \ \ \ \ {\rm (second\ solution)}
\label{sol2}
\eea
 where $f(p)$ is defined as
\bea
f(p)=S_2(p,\tilde{y})\,e^{2k\tilde{y}}
\eea
 and does not depend on $\tilde{y}$ because $S_2\propto e^{-2k\tilde{y}}$.
Now that $U_\nu+\alpha_\nu$ and the first solution for $S_1+\phi+W$ Eq.~(\ref{sol1}) are shown proportional to $e^{-2k\tilde{y}}$,
 we can perform a coordinate transformation Eq.~(\ref{trans}) to take the gauge where $U_\nu+\alpha_\nu=0$ and the first solution of $S_1+\phi+W$ vanishes.
Of the surviving fields, $\tau_{\mu\nu}$ is the massless graviton, and the combination of $S_1+\phi+W$ and $S_2$ specified by Eq.~(\ref{sol2}) describes the radion.
For clarity, below we present the metric that only includes the radion fluctuation:
\begin{align}
{\rm d}s^2&=\left[e^{-2k\tilde{y}}\eta_{\mu\nu}\left(1+f(\tilde{x})\right)-\frac{1}{k^2}\left(\frac{1}{2}-\frac{1}{d+1}e^{(d-1)k\tilde{y}}\right)\partial_\mu\partial_\nu f(\tilde{x})\right]{\rm d}\tilde{x}^\mu{\rm d}\tilde{x}^\nu-{\rm d}\tilde{y}^2
\label{radion}
\end{align}

\end{itemize}

It is now convenient to perform a coordinate transformation that (i) erases the derivative term in Eq.~(\ref{radion}), (ii) maintains the relation $G_{\mu 5}=0$,
 and (iii) keeps the positive boundary at the origin.
This is achieved by a coordinate transformation below,
\begin{align}
x^\mu&=\tilde{x}^\mu-\int^{\tilde{y}}_{-\infty}{\rm d}\tilde{y}^\prime\,\frac{1}{2k}\left(1-e^{(d-1)k\tilde{y}^\prime}\right)e^{2k\tilde{y}^\prime}\eta^{\mu\alpha}\partial_\alpha f(\tilde{x}),
\label{xtrans}\\
y&=\tilde{y}-\frac{1}{2k}\left(1-e^{(d-1)k\tilde{y}}\right)f(\tilde{x}).
\label{ytrans}
\end{align}
By neglecting terms of order $e^{2(d-1)kr}f(\tilde{x})^2$, the radion metric Eq.~(\ref{radion}) is re-expressed as
\begin{align}
{\rm d}s^2&=\exp\left[-2ky+e^{(d-1)ky}f(x)\right]\eta_{\mu\nu}{\rm d}x^\mu{\rm d}x^\nu
-\left(1-\frac{d-1}{2}e^{(d-1)ky}f(x)\right)^2{\rm d}y^2.
\label{radion2}
\end{align}
\\

\subsection{Step 2}

We situate the negative boundary in a way that the radion solution Eq.~(\ref{radion2}) obeys the equation of motion at the negative boundary.

To find the configuration of the negative boundary, suppose the negative boundary is given by
\bea
y+\zeta(x,y) \ = \ r\ \ (={\rm constant}).
\eea
We will constrain $\zeta$ by requiring that the equation of motion at the negative boundary be satisfied by the radion solution.
To facilitate calculation, we perform a coordinate transformation
\begin{align}
\hat{x}^\mu&=x^\mu+\int^{y}{\rm d}y'\,e^{2ky'}\eta^{\mu\alpha}\partial_\alpha \zeta(x,y'),
\label{xtrans2}\\
\hat{y}&=y+\zeta(x,y),
\label{ytrans2}
\end{align}
  so that the negative boundary is given by $\hat{y}=r$ and we still have $G_{\mu 5}=0$. 
The radion solution Eq.~(\ref{radion2}) is re-expressed as
\begin{align}
{\rm d}s^2=&\left[\exp\left[-2k\hat{y}+e^{(d-1)k\hat{y}}f(\hat{x})+2k\,\zeta(\hat{x},\hat{y})\right]\eta_{\mu\nu}
-2e^{-2k\hat{y}}\int^{\hat{y}}{\rm d}y'\,e^{2ky'}\,\partial_\mu\partial_\nu \zeta(\hat{x},y')\right]{\rm d}\hat{x}^\mu{\rm d}\hat{x}^\nu
\nn\\
&-\left(1-\frac{d-1}{2}e^{(d-1)k\hat{y}}f(\hat{x})-\frac{\partial \zeta(\hat{x},\hat{y})}{\partial \hat{y}}\right)^2 {\rm d}y^2.
\label{radion3}
\end{align}
In this coordinate, the equation of motion at the negative boundary that follows from Eq.~(\ref{variational3}) takes the form
\begin{align}
&\left.\frac{1}{\sqrt{-G_{55}}}\frac{1}{2}\frac{\partial g_{\mu\nu}}{\partial \hat{y}}-\frac{1}{\sqrt{-G_{55}}}\frac{1}{2}g_{\mu\nu}g^{\alpha\beta}\frac{\partial g_{\alpha\beta}}{\partial \hat{y}}-g_{\mu\nu}\ d\ k\right\vert_{\hat{y}=r}=0,
\label{beomp}
\end{align}

We require that the radion solution Eq.~(\ref{radion3}) obey the equation of motion~(\ref{beomp}).
Plugging Eq.~(\ref{radion3}) into Eq.~(\ref{beomp}) and working in the first order of $\zeta$ and $f$,
 we get
\begin{align}
\partial_\mu\partial_\nu\zeta(\hat{x},\hat{y}=r) \ = \ 0,
\label{constraint}
\end{align}
 which is regarded as a constraint on $\zeta$.
The only solution to Eq.~(\ref{constraint}) is $\zeta(\hat{x},\hat{y})=\zeta(\hat{y})$ if, as we assume,
 there is no preferred direction in 4D spacetime
 $c^\mu$ that gives a solution like $\zeta(\hat{x},\hat{y})=c^\mu \hat{x}_\mu$.
From $\zeta(\hat{x},\hat{y})=\zeta(\hat{y})$, we conclude that any $x$-dependent surface in the coordinate of Eqs.~(\ref{xtrans}),(\ref{ytrans})
 cannot be a configuration of the negative boundary,
 while all the $y=$(constant) surfaces are qualified to be such a configuration.
\\

To summarize, there exists a radion solution satisfying the linearized equations of motion in the bulk and at the two boundaries,
 given by
\begin{align}
{\rm d}s^2&=\exp\left[-2ky+e^{(d-1)ky}f(x)\right]\eta_{\mu\nu}{\rm d}x^\mu{\rm d}x^\nu
-\left(1-\frac{d-1}{2}e^{(d-1)ky}f(x)\right)^2{\rm d}y^2,
\label{radionfinal}
\end{align}
 provided both the boundaries are $y=$(constant) surfaces in the same coordinate.
\\

The kinetic term for the radion field $f(x)$ in 4D effective theory is derived as follows.
Plugging the metric Eq.~(\ref{radionfinal}) into the action Eq.~(\ref{action}), we get
\begin{align}
S_{\rm grav}=&M^d\int_0^r{\rm d}y\int{\rm d}^{d+1}x\ e^{-(d+1){\cal A}}\,\frac{{\cal A}'}{k}
\left\{-2(d+1)k^2-2d \ e^{2{\cal A}}\square {\cal A}+2\frac{e^{2{\cal A}}}{{\cal A}'}\square {\cal A}'\right.
\nn\\
&\left. \ \ \  \ \ \ \ \ \ \ \ \ \ \ \ \ \ \ \ \ \ \ \ \ \ \ \ \ \ \ \ \ \ \ +d(d-1)e^{2{\cal A}}\partial_\alpha {\cal A}\ \partial^\alpha {\cal A}-2(d-1)\frac{e^{2{\cal A}}}{{\cal A}'}\partial_\alpha {\cal A}\ \partial^\alpha {\cal A}'\right\}
\nn\\
&+M^d\left.\int{\rm d}^{d+1}x\ e^{-(d+1){\cal A}}\ 2k\right\vert_{y=0}
+M^d\left.\int{\rm d}^{d+1}x\ e^{-(d+1){\cal A}}(-2k)\right\vert_{y=r}
\nn\\
=&M^d\int_0^r{\rm d}y\int{\rm d}^{d+1}x\ \frac{\partial}{\partial y}\left[
2k\,e^{-(d+1){\cal A}}+2d\left(\frac{2}{-d+1}\right)^2\frac{1}{k}\ \frac{1}{2}\partial_\alpha\left(e^{\frac{-d+1}{2}{\cal A}}\right)
\partial^\alpha\left(e^{\frac{-d+1}{2}{\cal A}}\right)\right]
\nn\\
&+M^d\left.\int{\rm d}^{d+1}x\ e^{-(d+1){\cal A}}\ 2k\right\vert_{y=0}
+M^d\left.\int{\rm d}^{d+1}x\ e^{-(d+1){\cal A}}(-2k)\right\vert_{y=r}
\nn\\
=&M^d\int{\rm d}^{d+1}x\ \left[2d\left(\frac{2}{-d+1}\right)^2\frac{1}{k}\ \frac{1}{2}\partial_\alpha\left(e^{\frac{-d+1}{2}{\cal A}}\right)
\partial^\alpha\left(e^{\frac{-d+1}{2}{\cal A}}\right)\right]^{y=r}_{y=0}
\nn\\
=&2d\,\frac{M^d}{k}
\int{\rm d}^{d+1}x\ 
\frac{1}{4}\left(e^{(d-1)kr}\exp\left[\frac{d-1}{2}e^{(d-1)kr}f(x)\right]-\exp\left[\frac{d-1}{2}f(x)\right]\right)\frac{1}{2}\partial_\alpha f(x)\partial^\alpha f(x)
\end{align}
 where ${\cal A}=ky-\frac{1}{2}e^{(d-1)ky}f(x)$ and ${\cal A}'=\partial_y{\cal A}$. The term quadratic in $f(x)$ is extracted as
\bea
S_{\rm grav}\vert_{\rm quad}\ = \ 2d\,\frac{M^d}{k}\frac{e^{(d-1)kr}-1}{4}\int{\rm d}^{d+1}x \ \frac{1}{2}\partial_\alpha f(x)\partial^\alpha f(x),
\label{radionkin}
\eea
 in agreement with the result in the literature.
\\

\section{Radion Potential from a Bulk Gauge Field}

\subsection{Formula for the 1-loop Effective Potential}

We introduce a gauge field in the bulk and derive the 1-loop effective potential for the radion generated by its Casimir energy.
In the calculation, we replace the radion field $f(x)$ with an $x$-independent vacuum expectation value $\langle f\rangle$.
In this case, by a further coordinate transformation
\begin{align}
{\rm new} \ x \ &= \ x,
\label{xtrans3}\\
{\rm new} \ y \ &= \ y-\frac{1}{2k}e^{(d-1)ky}\langle f\rangle+\frac{1}{2k}, \ \ \ \ \ \ \langle f\rangle \ {\rm :vacuum \ expectation \ value \ of} \ f(x),
\label{ytrans3}
\end{align}
 we can erase the radion from the metric as
\begin{align}
 {\rm d}s^2=e^{-2ky}\eta_{\mu\nu}{\rm d}x^\mu{\rm d}x^\nu-{\rm d}y^2.
 \nn
\end{align}
This transformation leaves the positive boundary unchanged, but renders the position of the negative boundary $\langle f\rangle$-dependent, as
\begin{align}
{\rm positive \ boundary:} \ \ &y=0,
\nn\\
{\rm negative \ boundary:} \ \ &y=r-\frac{1}{2k}e^{(d-1)kr}\langle f\rangle+\frac{1}{2k},
\nn
\end{align}
 where $r$ is a constant that corresponds to the position of the negative boundary in the old coordinate.
In the rest of the paper, we use the new coordinate and express the radion VEV in terms of the boundary distance, $r_f$, given by
\bea
r_f \ = \ r-\frac{1}{2k}e^{(d-1)kr}\langle f\rangle+\frac{1}{2k}.
\label{rf}
\eea
\\

The action for the gauge field reads (we fix $d=3$ hereafter)
\footnote{
In this paper, we neglect boundary-localized kinetic terms.
}
\begin{align}
&S_{\rm gauge}=\int^{r_f}_0{\rm d}y\int{\rm d}^4x\ e^{-4ky}\left[-\frac{1}{4}F^a_{MN}F^{aMN}
-e^{4ky}\frac{1}{2\xi}\left\{\partial_\mu A^{a\mu}-\xi \, \partial_5(e^{-2ky}A_5^a)\right\}^2\right.
\nn\\
&\ \ \ \ \ \ \ \ \ \ \ \ \ \ \ \ \ \ \ \ \ \ \ \ \ \ \  \ \ \ \ \ \ \ \  \ \ \ \  \ \ \ \ \ \  \ \  \ \ \ \ \ \ \left.+e^{2ky}\,b^a\left\{\partial^\mu D_\mu^{ac}-\xi\,\partial_5\left(e^{-2ky}D_5^{ac}\right)\right\}c^c\right],
\label{gauge}
\end{align}
 where $\xi$ is a gauge-fixing parameter and $b,c$ are ghost fields.
Our gauge fixing procedure is the same as Ref.~\cite{Randall:2001gb}.
Note that $r_f$, and hence the radion VEV $\langle f\rangle$, enters into the end point of the $y$ integral.

To compute the 1-loop effective potential,
 we extract the quadratic terms from Eq.~(\ref{gauge}). We omit the gauge index hereafter.
The quadratic part is
\begin{align}
S_{\rm gauge}\vert_{\rm quad}&=\int^{r_f}_0{\rm d}y\int{\rm d}^4x\,\left[
\frac{1}{2}A_\mu\left(\eta^{\mu\nu}\square-\left(1-\frac{1}{\xi}\right)\partial^\mu\partial^\nu\right)A_\nu
-\frac{1}{2}A_\mu\partial_5 (e^{-2ky}(\partial_5A^{\mu})) \right.
\nn\\
&\ \ \ \ \ \ \ \ \ \ \ \ \left.
-\frac{1}{2}e^{-2ky}A_5\square A_5+\frac{\xi}{2}e^{-2ky}A_5\partial_5^2(e^{-2ky}A_5)
+e^{-2ky}b\square c-\xi\,e^{-2ky}b\ \partial_5(e^{-2ky}\partial_5 c) \right]
\label{quad}\\
& +\left[\int{\rm d}^4x\ e^{-2ky}(\partial^\mu A_\mu)A_5
+\frac{1}{2}A_\mu\,e^{-2ky}(\partial_5A^{\mu})
-\frac{\xi}{2}\,e^{-2ky} A_5\partial_5(e^{-2ky}A_5)
\right]^{y=r_f}_{y=0}.
\label{bcterms}
\end{align}

Applying variational principle to the boundary term Eq.~(\ref{bcterms}),
 we get either $(\partial_5A_\mu,\,A_5)=(0,0)$ or $(A_\mu,\,\partial_5(e^{-2ky}A_5))=(0,0)$
 at $y=0$ and $r_f$.
We choose {\bf Neumann-Neumann} condition for $A_\mu$,
\bea
(\partial_5A_\mu,\,A_5)\vert_{y=0,\,r_f}\,=\,(0,\,0).
\label{bc}
\eea
The boundary condition for the ghosts $b,c$ is not derived from variational principle, 
 but follows from Hermicity of the Lagrangian.
The $y$ derivatives of $c$ in Eq.~(\ref{quad}) are rewritten as
\begin{align}
\int_0^{r_f}{\rm d}y \ e^{-2ky}b\ \partial_5(e^{-2ky}\partial_5 c)
&=\int_0^{r_f}{\rm d}y \ \partial_5(e^{-2ky}\partial_5(e^{-2ky}b))\ c
\\
&+\left[\ e^{-2ky}b \ e^{-2ky}\partial_5c-\partial_5(e^{-2ky}b)e^{-2ky}c\ \right]^{y=r_f}_{y=0}.
\label{btghosts}
\end{align}
We require the boundary term Eq.~(\ref{btghosts}) to vanish, so that
 the Hermicity relation for the operator $\partial_5( e^{-2ky}\partial_5(\cdot))$ is obtained.
Then we get either $(\partial_5(e^{-2ky}b),\,\partial_5c)=(0,0)$ or $(e^{-2ky}b,\,c)=(0,0)$
 at $y=0$ and $r_f$.
Because the gauge fixing function $\partial_\mu A^\mu-\xi\partial_5(e^{-2ky}A_5)$ is non-vanishing at 
 $y=0$ and $r_f$ when Eq.~(\ref{bc}) is chosen, below is the correct boundary condition:
\begin{align}
\left(\,\partial_5(e^{-2ky}b),~\partial_5c\,\right)\vert_{y=0,\,r_f}\,=\,(0,\,0).
\label{bcghosts}
\end{align}

We expand $A_\mu,\,A_5,\,e^{-2ky}b,\,c$ into eigenfunctions of the operators in Eq.~(\ref{quad})
 obeying the boundary conditions Eqs.~(\ref{bc}),(\ref{bcghosts}).
Later when we calculate a summation over eigenvalues,
 it is vastly convenient that different eigenfunctions have different 4D momentum squared $p^2$
 so that the summation becomes a mere integral over $p^2$.
Therefore, we regard the coefficient of each $\square$ operator (1 for $A_\mu$, $e^{-2ky}$ for $A_5$, 1 for $e^{-2ky}b$ and $c$) as a weighting function
 and write the eigenvalue equations as
\begin{align}
&\left(\eta^{\mu\nu}\square-\left(1-\frac{1}{\xi}\right)\partial^\mu\partial^\nu\right)A_\nu-\partial_5(e^{-2ky}\,(\partial_5A^{\mu}))
=\lambda_{A_\mu}\,A^\mu,
\label{eigen1}\\
&-e^{-2ky}\square A_5+\xi \,e^{-2ky}\partial_5^2(e^{-2ky}\,A_5)=e^{-2ky}\lambda_{A_5}\,A_5,
\label{eigen2}\\
&\square c-\xi\,\partial_5(e^{-2ky}\,\partial_5 c)=\lambda_c\,c\ \ \ \ \ \ \ ({\rm the \ same \ for} \ e^{-2ky}b).
\label{eigen3}
\end{align}
Note the non-trivial weighting function $e^{-2ky}$ in front of $\lambda_{A_5}$
\footnote{
Due to this weighting function, the orthonormality relation for the eigenfunctions of Eq.~(\ref{eigen2}) becomes
\bea
\int_0^{r_f}{\rm d}y\, e^{-2ky}A_5^{(n)}(p,y)A_5^{(m)}(q,y)\propto\delta(p^2-q^2)\delta_{n,m}.
\eea
}.
The eigenvalues of Eqs.~(\ref{eigen1})-(\ref{eigen3}) under the boundary conditions Eqs.~(\ref{bc}),(\ref{bcghosts}) are~\cite{Davoudiasl:1999tf,Pomarol:1999ad}
\begin{align}
&\lambda_{A_\mu}^{(0,\pm\&L)}(p)=-p^2, \ \ \ \ \ \ \ \ \lambda_{A_\mu}^{(n,\pm\&L)}(p)=-p^2+x_n^2(ke^{-kr_f})^2
\nn\\
& \lambda_{A_\mu}^{(0,S)}(p)=-\frac{1}{\xi}p^2, \ \ \ \ \ \ \lambda_{A_\mu}^{(n,S)}(p)=-\frac{1}{\xi}p^2+x_n^2(ke^{-kr_f})^2,
\nn\\
&\lambda_{A_y}^{(n)}(p)=p^2-\xi\,x_n^2(ke^{-kr_f})^2,
\nn\\
&\lambda_c^{(0)}(p)=-p^2, \ \ \ \ \ \ \ \ \lambda_c^{(n)}(p)=-p^2+\xi\,x_n^2(ke^{-kr_f})^2,
\ \ \ \ \ \ \ \ \ \ \ \ \ \ n=1,2,3,...
\label{eigenvalues}
\end{align}
 where $p^\mu$ is a 4D momentum, and $x_n$ satisfies
\begin{align}
&J_0(x_n)Y_0(x_n e^{-kr_f})-Y_0(x_n)J_0(x_n e^{-kr_f})=0,\\
& \ \ \ \ \ \ \ ...>x_{n+1}>x_n>....>x_2>x_1>0.
 \nn
\end{align}
$\pm\&L$ refers to the transverse and longitudinal components and $S$ to the scalar component of $A_\mu$.

Finally, the 1-loop effective potential is obtained from the eigenvalues Eq.~(\ref{eigenvalues}) as
\begin{align}
&V_{\rm eff}(e^{-kr_f})-V_{\rm eff}(0)
\nn\\
&=-\frac{i}{2}n_{\rm g}\sum_{n=1}^\infty\int\frac{{\rm d}^Dp}{(2\pi)^D}\,
\left\{\,
\frac{D-1}{2}\log\left[\frac{-p^2+x_n^2(ke^{-kr_f})^2}{-p^2}\right]+\frac{1}{2}\log\left[\frac{-p^2/\xi+x_n^2(ke^{-kr_f})^2}{-p^2/\xi}\right]\right.
\nn\\
& \ \ \ \ \ \ \ \ \ \ \ \ \ \ \ \ \ \ \ \ \ \ \ \ \ \ \ \ \ \ \ \ \ \ 
\left.+\frac{1}{2}\log\left[\frac{-p^2+\xi\,x_n^2(ke^{-kr_f})^2}{-p^2}\right]-\log\left[\frac{-p^2+\xi\,x_n^2(ke^{-kr_f})^2}{-p^2}\right]\,\right\},
\label{veff}
\end{align}
 where $n_{\rm g}$ denotes the number of generators of the gauge group, and a general dimension $D$ is considered.
The first term originates from the transverse and scalar components of $A_\mu$, 
 the second term from the longitudinal component, the third term from $A_5$, and the fourth term from $e^{-2ky}b$ and $c$.
Obviously, the gauge dependence is cancelled.
 \\

 \subsection{Evaluation of the 1-loop Effective Potential}

\ \ \ \ \ First, we calculate the integral over 4D momentum $p$ with usual dimensional regularization with $D=4-2\epsilon$ and get
\begin{align}
V_{\rm eff}(e^{-kr_f})-V_{\rm eff}(0)=-\frac{1}{2}n_{\rm g}\frac{3-2\epsilon}{2}\frac{1}{(4\pi)^{2-\epsilon}}\Gamma(-2+\epsilon)
\left(ke^{-kr_f}\right)^{4-2\epsilon}\sum_{n=1}^\infty\,x_n^{4-2\epsilon}.
\label{veff2}
\end{align}
\\

Next, we discuss the regularization and renormalization of the infinite summation over $n$.

One na\"ively thinks that it is achieved by rewriting the summation for Re$(s)>1$, with the help of the integral expression of the Gamma function
 and the residue theorem, as
\footnote{
For large $n$, $x_n$ approaches to $\frac{n\pi}{1-e^{-kr_f}}$.
Hence, for $t\to+0$, $\sum_{n=1}^\infty e^{-x_n\,t}$ approaches to $\left(1-e^{t\pi/(1-e^{-kr_f})}\right)^{-1}$.
Therefore, $\sum_{n=1}^\infty e^{-x_n\,t}$ has an order-1 pole at $t=0$, and the integral over $t$ in Eq.~(\ref{wrong})
 converges at $t=0$.
}
\begin{align}
\sum_{n=1}^\infty\,x_n^{-s}&=\frac{1}{\Gamma(s)}\sum_{n=1}^\infty\left(\int_0^\infty{\rm d}t \ e^{-x_n\,t}t^{s-1}\right)
\nn\\
&=\frac{1}{\Gamma(s)}\int_0^\infty{\rm d}t\ t^{s-1}\left(\sum_{n=1}^\infty e^{-x_n\,t}\right)
\nn\\
&=\frac{1}{2\pi i}\frac{1}{\Gamma(s)}\int_0^\infty{\rm d}t\ t^{s-1}
\oint_C{\rm d}z \ e^{-z\,t} \ 
\frac{\frac{{\rm d}}{{\rm d}z}\left\{J_0(z)Y_0(ze^{-kr_f})-Y_0(z)J_0(ze^{-kr_f})\right\}}{J_0(z)Y_0(ze^{-kr_f})-Y_0(z)J_0(ze^{-kr_f})},
\ \ \ \ \ {\rm Re}(s)>1
\label{wrong}\\
&\ \ \ \ \ \ \ \ \ \ \ \ \ \ \ \ \  \ \ \ \ \ \ \ \ \ \ \ \ \ \ \ \ \ \ \ \  \ \ \ \ \ \ \  \ \ \ \ \ \ \ \ \ \ \ \ \  \ \ \ \ \ \ \ {\rm (we \ do \ not \ use \ this)}\nn
\end{align}
 where $C$ is a contour that encircles the whole real positive axis.
One then performs the analytic continuation of Eq.~(\ref{wrong}) to the following integral function defined for $s\neq1$:
\begin{align}
&\frac{1}{2\pi i}\frac{1}{\Gamma(s)}\frac{1}{e^{2\pi s\, i}-1}\oint_{C_{\rm keyhole}}{\rm d}w\ w^{s-1}
\oint_C{\rm d}z \ e^{-z\,w} \ 
\frac{\frac{{\rm d}}{{\rm d}z}\left\{J_0(z)Y_0(ze^{-kr_f})-Y_0(z)J_0(ze^{-kr_f})\right\}}{J_0(z)Y_0(ze^{-kr_f})-Y_0(z)J_0(ze^{-kr_f})}, \ \ \ \ \ s\neq1
\label{wrong2}\\
&\ \ \ \ \ \ \ \ \ \ \ \ \ \ \ \ \  \ \ \ \ \ \ \ \ \ \ \ \ \ \ \ \ \ \ \ \  \ \ \ \ \ \ \  \ \ \ \ \ \ \ \ \ \ \ \ \  \ \ \ \ \ \ \ \ \ \ \ \  \ \ \ \ \ \ \ {\rm (we \ do \ not \ use \ this)}\nn
\end{align}
  where the branch cut of $w$ is on the non-negative real axis, and $C_{\rm keyhole}$ is the contour of $w$ 
  defined as the
  $\rho\to+0$ limit of the contour that goes from $w=\infty+i0$ to $w=\rho+i0$, encircles the $w=0$ point with radius $\rho$, and goes from $\rho-i0$ to $\infty-i0$.
The value of Eq.~(\ref{wrong2}) at $s=-4+2\epsilon$ is regarded as the regularized and renormalized value of $\sum_{n=1}^\infty\,x_n^{4-2\epsilon}$.

A problem in the above procedure is that the function to be analytically continued, 

\noindent
$\oint_C{\rm d}z \ e^{-z\,t} \ 
\frac{\frac{{\rm d}}{{\rm d}z}\left\{J_0(z)Y_0(ze^{-kr_f})-Y_0(z)J_0(ze^{-kr_f})\right\}}{J_0(z)Y_0(ze^{-kr_f})-Y_0(z)J_0(ze^{-kr_f})}$,
depends on $e^{-kr_f}$. Thus, performing its analytic continuation is against the principle~(\ref{principle}) and leads to a wrong result, as expounded in Introduction.
\\

\begin{figure}[H]
\includegraphics[width=10cm]{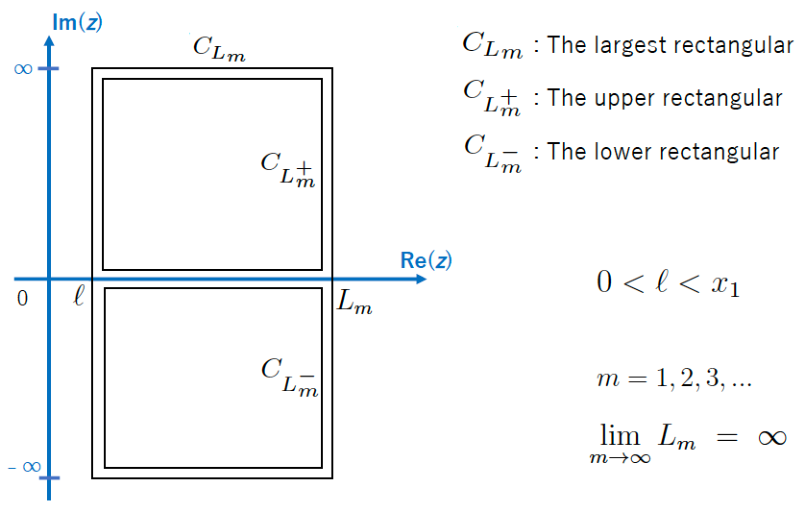}
\caption{
Contours on $z$ plane, $C_{L_m},\,C_{L_m^+},\,C_{L_m^-}$, that appear in Eq.~(\ref{integral}).
All the contours are counter-clockwise.
$\ell$ is an arbitrary positive number that is smaller than the smallest positive zero of $J_0(z)Y_0(ze^{-kr_f})-Y_0(z)J_0(ze^{-kr_f})$, which is $x_1$.
}
\label{contours}
\end{figure}

To avoid the problem, we propose a new regularization and renormalization procedure below.
We rewrite the summation for Re$(s)>1$ as (the branch cut of $z$ is on the non-positive real axis)
\begin{align}
\sum_{n=1}^\infty\,x_n^{-s}=\frac{1}{2\pi i}\lim_{m\to\infty}&\left[
\oint_{C_{L_m}}{\rm d}z\ z^{-s} \ \frac{\frac{{\rm d}}{{\rm d}z}\left\{J_0(z)Y_0(ze^{-kr_f})-Y_0(z)J_0(ze^{-kr_f})\right\}}{J_0(z)Y_0(ze^{-kr_f})-Y_0(z)J_0(ze^{-kr_f})}\right.
\nn\\
&+\oint_{C_{L_m^+}}{\rm d}z\ z^{-s}\left(\frac{H_1^{(2)}(z)}{H_0^{(2)}(z)}+e^{-kr_f}\frac{H_1^{(1)}(ze^{-kr_f})}{H_0^{(1)}(ze^{-kr_f})}\right)
\nn\\
&\left.+\oint_{C_{L_m^-}}{\rm d}z\ z^{-s}\left(\frac{H_1^{(1)}(z)}{H_0^{(1)}(z)}+e^{-kr_f}\frac{H_1^{(2)}(ze^{-kr_f})}{H_0^{(2)}(ze^{-kr_f})}\right)\right], \ \ \ \ \ \ \ {\rm Re}(s)>1
\label{integral}\\
& \ \ \ \ \ \ \ \ \ \ \ \ \ \ \ \ \ \ \ \ \ \ \ \ \ \ \ \ \ \ \ \ \ \ \ \ \ \ \ \ \ \ \ \ \ \ \ \ \ \ \ \ \ \ \ \ \ \ \ \ \ \ \ \ \ \ \ \ \ \ {\rm (we \ use \ this)} 
\nn
\end{align}
 where $H_{\nu}^{(1)}(z)$, $H_{\nu}^{(2)}(z)$ denote the Hankel functions of the first and second kind.
$C_{L_m},\,C_{L_m^+},\,C_{L_m^-}$ with $m=1,2,3,...$ are infinite sequence of contours
 which are depicted in Fig.~\ref{contours} for each $m$ and which satisfy $L_m\to\infty$ as $m\to\infty$.
Note that since $z^{-s}\frac{H_1^{(1)}(z)}{H_0^{(1)}(z)}$, $z^{-s}\frac{H_1^{(2)}(z)}{H_0^{(2)}(z)}$
 are regular for $z\neq0$, the second and third terms of Eq.~(\ref{integral}) give 0.
The essence of Eq.~(\ref{integral}) is that the function with non-factorizable $e^{-kr_f}$ dependence,
  $\frac{\frac{{\rm d}}{{\rm d}z}\left\{J_0(z)Y_0(ze^{-kr_f})-Y_0(z)J_0(ze^{-kr_f})\right\}}{J_0(z)Y_0(ze^{-kr_f})-Y_0(z)J_0(ze^{-kr_f})}$,
  is asymptoted by the functions with factorizable $e^{-kr_f}$ dependences, $\frac{H_1^{(i)}(z)}{H_0^{(i)}(z)}$ and $e^{-kr_f}\frac{H_1^{(i)}(ze^{-kr_f})}{H_0^{(i)}(ze^{-kr_f})}$.

Eq.~(\ref{integral}) is recast into
\begin{align}
&(\ref{integral})=\frac{1}{2\pi i}\lim_{m\to\infty}\ [
\nn\\
&\ \ \ \int^0_{-\infty}i{\rm d}v\ z^{-s} \left.\left[\frac{\frac{{\rm d}}{{\rm d}z}\left\{J_0(z)Y_0(ze^{-kr_f})-Y_0(z)J_0(ze^{-kr_f})\right\}}{J_0(z)Y_0(ze^{-kr_f})-Y_0(z)J_0(ze^{-kr_f})}
+\frac{H_1^{(1)}(z)}{H_0^{(1)}(z)}+e^{-kr_f}\frac{H_1^{(2)}(ze^{-kr_f})}{H_0^{(2)}(ze^{-kr_f})}\right]\right\vert_{z=L_m+iv}
\nn\\\label{L-}\\
&+\int^{\infty}_0i{\rm d}v\ z^{-s} \left.\left[\frac{\frac{{\rm d}}{{\rm d}z}\left\{J_0(z)Y_0(ze^{-kr_f})-Y_0(z)J_0(ze^{-kr_f})\right\}}{J_0(z)Y_0(ze^{-kr_f})-Y_0(z)J_0(ze^{-kr_f})}
+\frac{H_1^{(2)}(z)}{H_0^{(2)}(z)}+e^{-kr_f}\frac{H_1^{(1)}(ze^{-kr_f})}{H_0^{(1)}(ze^{-kr_f})}\right]\right\vert_{z=L_m+iv}
\nn\\\label{L+}\\
&+\int^\ell_{L_m}{\rm d}u\ z^{-s} \left.\left[\frac{\frac{{\rm d}}{{\rm d}z}\left\{J_0(z)Y_0(ze^{-kr_f})-Y_0(z)J_0(ze^{-kr_f})\right\}}{J_0(z)Y_0(ze^{-kr_f})-Y_0(z)J_0(ze^{-kr_f})}
+\frac{H_1^{(2)}(z)}{H_0^{(2)}(z)}+e^{-kr_f}\frac{H_1^{(1)}(ze^{-kr_f})}{H_0^{(1)}(ze^{-kr_f})}\right]\right\vert_{z=u+i\infty}
\nn\\\label{upper}\\
&+\int_{\infty}^0i{\rm d}v\ z^{-s} \left.\left[\frac{\frac{{\rm d}}{{\rm d}z}\left\{J_0(z)Y_0(ze^{-kr_f})-Y_0(z)J_0(ze^{-kr_f})\right\}}{J_0(z)Y_0(ze^{-kr_f})-Y_0(z)J_0(ze^{-kr_f})}
+\frac{H_1^{(2)}(z)}{H_0^{(2)}(z)}+e^{-kr_f}\frac{H_1^{(1)}(ze^{-kr_f})}{H_0^{(1)}(ze^{-kr_f})}\right]\right\vert_{z=\ell+iv}
\nn\\\label{l+}\\
&+\int_0^{-\infty}i{\rm d}v\ z^{-s} \left.\left[\frac{\frac{{\rm d}}{{\rm d}z}\left\{J_0(z)Y_0(ze^{-kr_f})-Y_0(z)J_0(ze^{-kr_f})\right\}}{J_0(z)Y_0(ze^{-kr_f})-Y_0(z)J_0(ze^{-kr_f})}
+\frac{H_1^{(1)}(z)}{H_0^{(1)}(z)}+e^{-kr_f}\frac{H_1^{(2)}(ze^{-kr_f})}{H_0^{(2)}(ze^{-kr_f})}\right]\right\vert_{z=\ell+iv}
\nn\\\label{l-}\\
&+\int_\ell^{L_m}{\rm d}u\ z^{-s} \left.\left[\frac{\frac{{\rm d}}{{\rm d}z}\left\{J_0(z)Y_0(ze^{-kr_f})-Y_0(z)J_0(ze^{-kr_f})\right\}}{J_0(z)Y_0(ze^{-kr_f})-Y_0(z)J_0(ze^{-kr_f})}
+\frac{H_1^{(1)}(z)}{H_0^{(1)}(z)}+e^{-kr_f}\frac{H_1^{(2)}(ze^{-kr_f})}{H_0^{(2)}(ze^{-kr_f})}\right]\right\vert_{z=u-i\infty}
\nn\\\label{lower}\\
&+\int_\ell^{L_m}{\rm d}u\ u^{-s} \left[
\frac{H_1^{(2)}(u)}{H_0^{(2)}(u)}-\frac{H_1^{(1)}(u)}{H_0^{(1)}(u)}
+e^{-kr_f}\frac{H_1^{(1)}(ue^{-kr_f})}{H_0^{(1)}(ue^{-kr_f})}
-e^{-kr_f}\frac{H_1^{(2)}(ue^{-kr_f})}{H_0^{(2)}(ue^{-kr_f})}\right]
\label{real}
\\
].
\nn
\end{align}
The sum of the integrals Eqs.~(\ref{L-}),(\ref{L+}) is real and oscillates about 0 as $L_m$ increases.
Now \textit{we choose} $L_m$ \textit{such that}
\begin{align}
&(\ref{L+})+(\ref{L-})=0 \ \ {\rm for \ every} \ \ L_m \ \ \ \ \ (m=1,2,3,...),
\nn\\
&{\rm with} \ \ ...>L_{m+1}>L_m>......>L_2>L_1>0. \ \ \ \ \ \ \ \ \ \ \ \ \ 
\label{lmdef}
\end{align}
The above choice correctly gives $\lim_{m\to\infty}L_m=\infty$, because for large $m$ the sum of Eqs.~(\ref{L-}),(\ref{L+}) oscillates with period $2\pi/(1-e^{-kr_f})$ and the sequence $L_m,L_{m+1},L_{m+2},...$ becomes equally spaced.

Below we examine the rest of the integrals.

As shown in Appendix~A, the integrand of Eq.~(\ref{upper}) dissipates as $\propto e^{-2{\rm Im}(z)(1-e^{-kr_f})}$
 and that of Eq.~(\ref{lower}) dissipates as $\propto e^{2{\rm Im}(z)(1-e^{-kr_f})}$, uniformly with respect to Re$(z)$.
Hence, Eqs.~(\ref{upper}),(\ref{lower}) vanish for any $s$.

Each of the integrals Eqs.~(\ref{l+}),(\ref{l-}) is finite for any $s$ by the same dissipation rule as above.
Thus, we can take $s=-4+2\epsilon$ in Eqs.~(\ref{l+}),(\ref{l-}) without analytic continuation.

The last integral Eq.~(\ref{real}) is convergent only for Re$(s)>1$.
Its analytic continuation must be performed as follows:
Rewrite Eq.~(\ref{real}) as
\begin{align}
(\ref{real})=&\int_\ell^{L_m}{\rm d}u\ u^{-s} \left[
\frac{H_1^{(2)}(u)}{H_0^{(2)}(u)}-\frac{H_1^{(1)}(u)}{H_0^{(1)}(u)}\right]
\label{real1}\\
&+\int_\ell^{e^{-kr_f}}{\rm d}u\ u^{-s} \left[
e^{-kr_f}\frac{H_1^{(1)}(e^{-kr_f}u)}{H_0^{(1)}(e^{-kr_f}u)}-e^{-kr_f}\frac{H_1^{(2)}(e^{-kr_f}u)}{H_0^{(2)}(e^{-kr_f}u)}\right]
\label{real2}\\
&+\int_{e^{-kr_f}}^{L_m}{\rm d}u\ u^{-s} \left[
e^{-kr_f}\frac{H_1^{(1)}(e^{-kr_f}u)}{H_0^{(1)}(e^{-kr_f}u)}-e^{-kr_f}\frac{H_1^{(2)}(e^{-kr_f}u)}{H_0^{(2)}(e^{-kr_f}u)}\right].
\label{real3}
\end{align}
Eq.~(\ref{real3}) is further rewritten through a variable change $u\to e^{-kr_f}u$, as
\begin{align}
(\ref{real3})=&(e^{kr_f})^{-s}
\int_1^{e^{kr_f}L_m}{\rm d}u\ u^{-s} \left[
\frac{H_1^{(1)}(u)}{H_0^{(1)}(u)}-\frac{H_1^{(2)}(u)}{H_0^{(2)}(u)}\right].
\label{real4}
\end{align}
A crucial fact is that in the limit with $m\to\infty$, the integrals
\bea
\lim_{m\to\infty}\int_\ell^{L_m}{\rm d}u\ u^{-s} \left[
\frac{H_1^{(2)}(u)}{H_0^{(2)}(u)}-\frac{H_1^{(1)}(u)}{H_0^{(1)}(u)}\right], \ \ \ \ \ 
\lim_{m\to\infty}\int_1^{e^{kr_f}L_m}{\rm d}u\ u^{-s} \left[
\frac{H_1^{(1)}(u)}{H_0^{(1)}(u)}-\frac{H_1^{(2)}(u)}{H_0^{(2)}(u)}\right]
\nn
\eea
 are independent of $e^{-kr_f}$.
Therefore, the analytic continuations of $\lim_{m\to\infty}$(\ref{real1}), $\lim_{m\to\infty}$(\ref{real4})
 are performed independently of $e^{-kr_f}$.
Since $\lim_{m\to\infty}$(\ref{real1}) is totally independent of $e^{-kr_f}$, it is renormalized with the negative boundary tension in Eq.~(\ref{action}).
Since $\lim_{m\to\infty}$(\ref{real4}) is proportional to $(e^{kr_f})^{4-2\epsilon}$, it is renormalized with the positive boundary tension in Eq.~(\ref{action})
\footnote{
Remind the factor $(e^{-kr_f})^{4-2\epsilon}$ in Eq.~(\ref{veff2}) that multiplies $\sum_{n=1}^\infty\,x_n^{4-2\epsilon}$.
}.
Note that, to be consistent with the dimensional regularization of 4D momentum integral,
  the negative boundary tension must be proportional to $(e^{-kr_f})^{4-2\epsilon}$.
Eq.~(\ref{real2}) is convergent for any $s$ and needs no analytic continuation.

To summarize, we have performed the analytic continuation of $\sum_{n=1}^\infty\,x_n^{-s}$ from Re$(s)>1$
 to $s=-4+2\epsilon$ independently of $e^{-kr_f}$, and have shown that the divergent terms $\lim_{m\to\infty}$(\ref{real1}), $\lim_{m\to\infty}$(\ref{real4}) are respectively renormalized with the negative and positive boundary tensions.
The finite terms come from Eqs.~(\ref{l+}),(\ref{l-}),(\ref{real2}).

When $s=-4+2\epsilon$, the integrands of Eqs.~(\ref{l+}),(\ref{l-}),(\ref{real2}) are regular for $\{z\,|\,0\leq{\rm Re}(z) < x_1\}$.
Hence, by Cauchy theorem, the sum of Eqs.~(\ref{l+}),(\ref{l-}),(\ref{real2}) is identical for any $\ell$ in the range $x_1>\ell\geq0$.
Once we take $\ell=0$, Eq.~(\ref{real2}) becomes proportional to $(e^{kr_f})^{4-2\epsilon}$ and can be renormalized
 with the positive boundary tension (finite renormalization).
Therefore, to evaluate the finite terms, it suffices to calculate Eqs.~(\ref{l+}),(\ref{l-}) by setting $\ell=0$.
\\

Now we evaluate the full radion potential Eq.~(\ref{veff2}).

We isolate the $\frac{1}{\epsilon}$ poles by expanding the sum of Eqs.~(\ref{l+}),(\ref{l-}) (with $s=-4+2\epsilon$ and $\ell=0$) in terms of $\epsilon$.
In doing so, we note that since the function 
$z^4\frac{\frac{{\rm d}}{{\rm d}z}\left\{J_0(z)Y_0(ze^{-kr_f})-Y_0(z)J_0(ze^{-kr_f})\right\}}{J_0(z)Y_0(ze^{-kr_f})-Y_0(z)J_0(ze^{-kr_f})}$ is an odd function of $z$, its integral in the range $(-i\infty,i\infty)$ vanishes.
Therefore, the sum is expanded as
 \begin{align}
&(\ref{l+})+(\ref{l-})
 \nn\\
 &\left. = \int_{\infty}^0 i{\rm d}v\ z^4\left[\frac{H_1^{(2)}(z)}{H_0^{(2)}(z)}+\frac{H_1^{(1)}(-z)}{H_0^{(1)}(-z)}\right] \right\vert_{z=iv}
\label{61+62-a}
\\
 &\left. \ \ +\int_0^{-\infty} i{\rm d}v\ z^4\left[e^{-kr_f}\frac{H_1^{(1)}(z\,e^{-kr_f})}{H_0^{(1)}(z\,e^{-kr_f})}+e^{-kr_f}\frac{H_1^{(2)}(-z\,e^{-kr_f})}{H_0^{(2)}(-z\,e^{-kr_f})}\right] \right\vert_{z=iv}
\label{61+62-a2}
\\
& \ \ -2\epsilon
\int_{\infty}^0i{\rm d}v\ z^4 \log z
\left[\frac{\frac{{\rm d}}{{\rm d}z}\left\{J_0(z)Y_0(ze^{-kr_f})-Y_0(z)J_0(ze^{-kr_f})\right\}}
{J_0(z)Y_0(z\,e^{-kr_f})-Y_0(z)J_0(z\,e^{-kr_f})}\right. 
\nn\\
&\left.\left.  \ \ \ \ \ \ \ \ \ \ \ \ \ \ \ \ \ \  \ \ \ \ \  \ \ \ \ \ \ \ \ \ \ \ \ \ \ \ \ \ \  \ \ \ \ \  \ \ \ \ \ \ \ \ \ \ \ \ \ \ \ \ \ \ +\frac{H_1^{(2)}(z)}{H_0^{(2)}(z)}+e^{-kr_f}\frac{H_1^{(1)}(z\,e^{-kr_f})}{H_0^{(1)}(z\,e^{-kr_f})}\right]\right\vert_{z=iv}
\label{61+62-b}
\\
&\ \ -2\epsilon\int_0^{-\infty}i{\rm d}v\ z^4 \log z \left[\frac{\frac{{\rm d}}{{\rm d}z}\left\{J_0(z)Y_0(ze^{-kr_f})-Y_0(z)J_0(ze^{-kr_f})\right\}}
{J_0(z)Y_0(z\,e^{-kr_f})-Y_0(z)J_0(z\,e^{-kr_f})}\right.
\nn\\
&\left.\left. \ \ \ \ \ \ \ \ \ \ \ \ \ \ \ \ \ \  \ \ \ \ \  \ \ \ \ \ \ \ \ \ \ \ \ \ \ \ \ \ \  \ \ \ \ \  \ \ \ \ \ \ \ \ \ \ \ \ \ \ \ \ \ \ 
+\frac{H_1^{(1)}(z)}{H_0^{(1)}(z)}+e^{-kr_f}\frac{H_1^{(2)}(z\,e^{-kr_f})}{H_0^{(2)}(z\,e^{-kr_f})}\right]\right\vert_{z=iv}
\label{61+62-c}
\\
&+O(\epsilon^2).
\nn
\end{align}
Eq.~(\ref{61+62-a}) itself is finite and gives a $\frac{1}{\epsilon}$ pole in the radion potential.
Since Eq.~(\ref{61+62-a}) has no $e^{kr_f}$ dependence, this pole is renormalized with with the negative boundary tension.
Similarly, Eq.~(\ref{61+62-a2}) itself is finite and gives a $\frac{1}{\epsilon}$ pole in the radion potential.
Eq.~(\ref{61+62-a2}) is proportional to $(e^{kr_f})^4$ and so this pole is renormalized with with the positive boundary tension.
Eqs.~(\ref{61+62-b}),(\ref{61+62-c}) provide finite contributions to the radion potential.

Plugging Eqs.~(\ref{61+62-b}),(\ref{61+62-c}) into Eq.~(\ref{veff2}),
 we find that the renormalized radion potential is given by
\begin{align}
&\left.V_{\rm eff}(e^{-kr_f})-V_{\rm eff}(0)\right\vert_{\rm renormalized}
\nn\\
&=A+B(ke^{-kr_f})^4
\nn\\
&
-\frac{3n_{\rm g}}{128\pi^3}
\left[
\int^{\infty}_0{\rm d}v\ z^4 \log z
\left[
\frac{\frac{{\rm d}}{{\rm d}z}\left\{J_0(z)Y_0(ze^{-kr_f})-Y_0(z)J_0(ze^{-kr_f})\right\}}
{J_0(z)Y_0(z\,e^{-kr_f})-Y_0(z)J_0(z\,e^{-kr_f})}\right. \right.
\nn\\
&\left.\left.  \ \ \ \ \ \ \ \ \ \ \ \ \ \ \ \ \ \  \ \ \ \ \  \ \ \ \ \ \ \ \ \ \ \ \ \ \ \ \ \ \  \ \ \ \ \  \ \ \ \ \ \ \ \ \ \ \ \ \ \ \ \ \ \ +\frac{H_1^{(2)}(z)}{H_0^{(2)}(z)}+e^{-kr_f}\frac{H_1^{(1)}(z\,e^{-kr_f})}{H_0^{(1)}(z\,e^{-kr_f})}\right]\right\vert_{z=iv}
\nn\\
&\ \ \ \ \ \ \ \ \ \ +\int^0_{-\infty}{\rm d}v\ z^4 \log z 
\left[
\frac{\frac{{\rm d}}{{\rm d}z}\left\{J_0(z)Y_0(ze^{-kr_f})-Y_0(z)J_0(ze^{-kr_f})\right\}}
{J_0(z)Y_0(z\,e^{-kr_f})-Y_0(z)J_0(z\,e^{-kr_f})}\right.
\nn\\
&\left.\left.\left. \ \ \ \ \ \ \ \ \ \ \ \ \ \ \ \ \ \  \ \ \ \ \  \ \ \ \ \ \ \ \ \ \ \ \ \ \ \ \ \ \  \ \ \ \ \  \ \ \ \ \ \ \ \ \ \ \ \ \ \ \ \ \ \ 
+\frac{H_1^{(1)}(z)}{H_0^{(1)}(z)}+e^{-kr_f}\frac{H_1^{(2)}(z\,e^{-kr_f})}{H_0^{(2)}(z\,e^{-kr_f})}\right]\right\vert_{z=iv} \right]
\label{veff3}
\end{align}
 where $A,B$ are renormalization constants.
We numerically calculate the integrals in Eq.~(\ref{veff3}) and fit them as a function of $kr_f$, and obtain the following estimate:
\begin{align}
&\left.V_{\rm eff}(e^{-kr_f})-V_{\rm eff}(0)\right\vert_{\rm renormalized}
\ =\ A+B(ke^{-kr_f})^4-\frac{3n_{\rm g}}{128\pi^3}(ke^{-kr_f})^4 \frac{17.0}{(k r_f)^{1.12}}.
\label{veff4}
\end{align}
Eq.~(\ref{veff4}) contains a term proportional to $1/(k\,r_f)^{1.12}$, i.e. it has logarithmic dependence on the warp factor $e^{-kr_f}$.
This is in accord with the result of Ref.~\cite{Garriga:2002vf}.
However, the exact expression of the renormalized potential Eq.~(\ref{veff3}) is different from that work.
The logarithmic dependence on $e^{-kr_f}$ makes us hope that the radion potential could generate a large hierarchy from a natural value of $B$.
Unfortunately, however, this potential has no minimum for any $A,B$ and always destabilizes the radion VEV.

In the next subsection, we study whether the radion stabilization is possible when a boundary condition other than Eq.~(\ref{bc}) is selected.
\\

\subsection{Consequences of Other Boundary Conditions}

 \subsubsection{Dirichlet-Neumann Condition for $A_\mu$}
 
In place of Eq.~(\ref{bc}), we choose 
\bea
(A_\mu,\,\partial_5(e^{-2ky}A_5))\vert_{y=0}\,=\,(0,\,0) \ \ \ \ {\rm and} \ \ \ \ (\partial_5A_\mu,\,A_5)\vert_{y=r_f}\,=\,(0,\,0).
\label{bc2}
\eea
The radion potential is then proportional to
 \bea
 \sum_{n=1}^\infty x_n^{4-2\epsilon} \ \ \ \ \ \ {\rm with} \ \ \ J_1(x_n)Y_0(x_n e^{-kr_f})-Y_1(x_n)J_0(x_n e^{-kr_f})=0,
\eea
 which is regularized and renormalized by writing
\begin{align}
\sum_{n=1}^\infty\,x_n^{-s}=\frac{1}{2\pi i}\lim_{m\to\infty}&\left[
\oint_{C_{L_m}}{\rm d}z\ z^{-s} \ \frac{\frac{{\rm d}}{{\rm d}z}\left\{J_1(z)Y_0(ze^{-kr_f})-Y_1(z)J_0(ze^{-kr_f})\right\}}{J_1(z)Y_0(ze^{-kr_f})-Y_1(z)J_0(ze^{-kr_f})}\right.
\nn\\
&+\oint_{C_{L_m^+}}{\rm d}z\ z^{-s}\left(-\frac{1}{2}\frac{H_0^{(2)}(z)}{H_1^{(2)}(z)}+\frac{1}{2}\frac{H_2^{(2)}(z)}{H_1^{(2)}(z)}
+e^{-kr_f}\frac{H_1^{(1)}(ze^{-kr_f})}{H_0^{(1)}(ze^{-kr_f})}\right)
\nn\\
&\left.+\oint_{C_{L_m^-}}{\rm d}z\ z^{-s}\left(-\frac{1}{2}\frac{H_0^{(1)}(z)}{H_1^{(1)}(z)}+\frac{1}{2}\frac{H_2^{(1)}(z)}{H_1^{(1)}(z)}
+e^{-kr_f}\frac{H_1^{(2)}(ze^{-kr_f})}{H_0^{(2)}(ze^{-kr_f})}
\right)\right],
\label{integral2}
\\
 &\ \ \ \  \ \ \ \  \ \ \ \  \ \ \ \  \ \ \ \  \ \ \ \  \ \ \ \ \ \ \ \  \ \ \ \  \ \ \ \  \ \ \ \  \ \ \ \  \ \ \ \  {\rm Re}(s)>1
 \nn
\end{align}
 and taking the same procedures as the previous subsection.
The radion potential is estimated as
\begin{align}
\left.V_{\rm eff}(e^{-kr_f})-V_{\rm eff}(0)\right\vert_{\rm renormalized} \ &= \
A+B(ke^{-kr_f})^4+\frac{3n_{\rm g}}{128\pi^3}(ke^{-kr_f})^4 \frac{28.6}{(k r_f)^{1.02}}
\label{veff-dn}
\end{align}
 where $A,B$ are renormalization constants.
This potential has a minimum and can stabilize the radion VEV. Also, it has logarithmic dependence on $e^{-kr_f}$ and can generate a large hierarchy of $e^{-kr_f}$ from a natural value of $B$.
As an example, we impose the following renormalization condition to have the radion VEV stabilized at $kr_f=31\ (\simeq \log(10~{\rm TeV}/0.1M_P))$ and also have vanishing cosmological constant in 4D effective theory:
\begin{align*}
\left.\frac{{\rm d}}{{\rm d}r_f}V_{\rm eff}(e^{-kr_f})\right\vert_{\rm renormalized}\ = \ 
0 \ \ \ &{\rm and} \ \ \ 
\left.V_{\rm eff}(e^{-kr_f})\right\vert_{\rm renormalized}\ = \ 
0
\ \ \ {\rm at} \ \ \ k\,r_f=31.
\end{align*}
The value of $B$ fixed by the above renormalization condition is $B=-(0.16)^4n_{\rm g}$.
Although the value of $B$ has no solid physical meaning, we see that it is a natural value of $O((0.1)^4)$,
 which means that we have successfully generated a large hierarchy of $e^{-kr_f}$ without fine-tuning of $B$.

We study the mass of the radion.
Noting that the kinetic term of $f(x)$ is non-canonical as in Eq.~(\ref{radionkin}),
 we calculate the radion mass as
\begin{align}
m_{\rm radion}^2&=\frac{1}{2\cdot 3}\frac{k}{M^3}\frac{4}{e^{(3-1)kr}-1}\frac{{\rm d}^2}{{\rm d}\langle f\rangle^2}V_{\rm eff}(e^{-kr_f})
\\
&\simeq n_{\rm g}\left(\frac{k}{M}\right)^3 k^2e^{-2kr_f}\left\{\frac{0.0445}{(kr_f)^{3.02}}+\frac{0.0882}{(kr_f)^{2.02}}\right\}
\end{align}
 for $e^{-kr_f}\ll1$.
\\

 \subsubsection{Dirichlet-Dirichlet Condition for $A_\mu$}
In place of Eq.~(\ref{bc}), we choose
\bea
(A_\mu,\,\partial_5(e^{-2ky}A_5))\vert_{y=0,r_f}\,=\,(0,\,0).
\label{bc3}
\eea
The radion effective potential is then proportional to
 \bea
 \sum_{n=1}^\infty x_n^{4-2\epsilon} \ \ \ \ \ \ {\rm with} \ \ \ J_1(x_n)Y_1(x_n e^{-kr_f})-Y_1(x_n)J_1(x_n e^{-kr_f})=0,
\eea
 which is regularized and renormalized by writing
\begin{align}
\sum_{n=1}^\infty\,x_n^{-s}=\frac{1}{2\pi i}\lim_{m\to\infty}&\left[
\oint_{C_{L_m}}{\rm d}z\ z^{-s} \ \frac{\frac{{\rm d}}{{\rm d}z}\left\{J_1(z)Y_1(ze^{-kr_f})-Y_1(z)J_1(ze^{-kr_f})\right\}}{J_1(z)Y_1(ze^{-kr_f})-Y_1(z)J_1(ze^{-kr_f})}\right.
\nn\\
&+\oint_{C_{L_m^+}}{\rm d}z\ z^{-s}\left(\frac{H_2^{(2)}(z)}{H_1^{(2)}(z)}
-e^{-kr_f}\frac{H_0^{(1)}(ze^{-kr_f})}{H_1^{(1)}(ze^{-kr_f})}\right)
\nn\\
&\left.+\oint_{C_{L_m^-}}{\rm d}z\ z^{-s}\left(\frac{H_2^{(1)}(z)}{H_1^{(1)}(z)}
-e^{-kr_f}\frac{H_0^{(2)}(ze^{-kr_f})}{H_1^{(2)}(ze^{-kr_f})}
\right)\right], \ \ \ \ \ {\rm Re}(s)>1
\label{integral3}
\end{align}
 and taking the same procedure as the previous subsection.
We find that Eq.~(\ref{integral3}) is constant and does no depend on $e^{-kr_f}$ when $e^{-kr_f}\ll1$.
Therefore, the radion effective potential is simply given by
\begin{align}
\left.V_{\rm eff}(e^{-kr_f})-V_{\rm eff}(0)\right\vert_{\rm renormalized} \ &= \
A+B(ke^{-kr_f})^4
\label{veff-dd}
\end{align}
 where $A,B$ are renormalization constants.
Since the potential has exponential dependence on $k\,r_f$, we cannot generate a large hierarchy of $e^{-kr_f}\sim 10$~TeV$/(0.1M_P)$
 unless we fine-tune $B$ to an extremely small value.
However, the fine-tuning of $B$ undermines the motivation of RS-1 model and so this case is not theoretically interesting.
\\

\subsubsection{Neumann-Dirichlet Condition for $A_\mu$}
 
In place of Eq.~(\ref{bc}), we choose
\bea
(\partial_5A_\mu,\,A_5)\vert_{y=0}\,=\,(0,\,0) \ \ \ \ {\rm and} \ \ \ \ (A_\mu,\,\partial_5(e^{-2ky}A_5))\vert_{y=r_f}\,=\,(0,\,0).
\label{bc4}
\eea
The radion effective potential is then proportional to
 \bea
 \sum_{n=1}^\infty x_n^{4-2\epsilon} \ \ \ \ \ \ {\rm with} \ \ \ J_0(x_n)Y_1(x_n e^{-kr_f})-Y_0(x_n)J_1(x_n e^{-kr_f})=0,
\eea
 which is regularized and renormalized by writing
\begin{align}
\sum_{n=1}^\infty\,x_n^{-s}=\frac{1}{2\pi i}\lim_{m\to\infty}&\left[
\oint_{C_{L_m}}{\rm d}z\ z^{-s} \ \frac{\frac{{\rm d}}{{\rm d}z}\left\{J_0(z)Y_1(ze^{-kr_f})-Y_0(z)J_1(ze^{-kr_f})\right\}}{J_0(z)Y_1(ze^{-kr_f})-Y_0(z)J_1(ze^{-kr_f})}\right.
\nn\\
&+\oint_{C_{L_m^+}}{\rm d}z\ z^{-s}\left(\frac{H_1^{(2)}(z)}{H_0^{(2)}(z)}+\frac{e^{-kr_f}}{2}\frac{H_2^{(1)}(ze^{-kr_f})}{H_1^{(1)}(ze^{-kr_f})}
-\frac{e^{-kr_f}}{2}\frac{H_0^{(1)}(ze^{-kr_f})}{H_1^{(1)}(ze^{-kr_f})}\right)
\nn\\
&\left.+\oint_{C_{L_m^-}}{\rm d}z\ z^{-s}\left(\frac{H_1^{(1)}(z)}{H_0^{(1)}(z)}+\frac{e^{-kr_f}}{2}\frac{H_2^{(2)}(ze^{-kr_f})}{H_1^{(2)}(ze^{-kr_f})}
-\frac{e^{-kr_f}}{2}\frac{H_0^{(2)}(ze^{-kr_f})}{H_1^{(2)}(ze^{-kr_f})}
\right)\right],
\label{integral4}
\\
 &\ \ \ \  \ \ \ \  \ \ \ \  \ \ \ \  \ \ \ \  \ \ \ \  \ \ \ \ \ \ \ \  \ \ \ \  \ \ \ \  \ \ \ \  \ \ \ \  \ \ \ \  {\rm Re}(s)>1
 \nn
\end{align}
 and taking the same procedure as the previous subsection.
As in the case of Dirichlet-Dirichlet condition, 
 we find that Eq.~(\ref{integral4}) is constant and does no depend on $e^{-kr_f}$ when $e^{-kr_f}\ll1$,
 and hence the radion effective potential is given by
\begin{align}
\left.V_{\rm eff}(e^{-kr_f})-V_{\rm eff}(0)\right\vert_{\rm renormalized} \ &= \
A+B(ke^{-kr_f})^4
\label{veff-dd}
\end{align}
 where $A,B$ are renormalization constants.
Since the potential has exponential dependence on $k\,r_f$, we cannot generate a large hierarchy of $e^{-kr_f}\sim 10$~TeV$/(0.1M_P)$
 unless we fine-tune $B$ to an extremely small value.
Such a fine-tuning is against the motivation of RS-1 model and so the case of Neumann-Dirichlet condition is theoretically unappealing.
\\

\section{Conclusion}

We have studied the stabilization of the radion in Randall-Sundrum-1 model
 by the Casimir energy of a bulk gauge field.
We have performed the correct evaluation of the 1-loop effective potential of the radion generated by the Casimir energy, by noting that, for the correct regularization and renormalization, analytic continuation must be performed only on functions that are independent of the radion vacuum expectation value.
From the above calculation, we have found that a bulk gauge field satisfying Dirichlet condition at the positive (UV) boundary and Neumann condition at the negative (IR) boundary gives rise to an appropriate potential that stabilizes the radion vacuum expectation value in a way that a large hierarchy of the warp factor is generated without fine-tuning of the relevant renormalization constant, thus solving the big hierarchy problem.

\section*{Acknowledgement}
This work is partially supported by Scientific Grants by the Ministry
of Education, Culture, Sports, Science and Technology of Japan
(Nos.~16H00871, 16H02189, 17K05415 and 18H04590).

\section*{Appendix~A}

The Hankel functions of the first and second kind, $H_n^{(1)}(z)$, $H_n^{(2)}(z)$, are expressed as
\bea
H_n^{(1)}(z)=2\frac{z^n}{i\sqrt{\pi}\Gamma(n+\frac{1}{2})2^n}F_n(z), \ \ \ \ \
H_n^{(2)}(z) = 2\frac{z^n}{i\sqrt{\pi}\Gamma(n+\frac{1}{2})2^n}F_n(-z)
\label{hankelappendix}
\eea
where
\bea
F_n(z)=\frac{e^{iz}}{z}\int_0^\infty{\rm d}t\ e^{-t}\left(-2i\frac{t}{z}+\frac{t^2}{z^2}\right)^{n-\frac{1}{2}}.
\label{fdef}
\eea
The Bessel functions are expressed as
\bea
J_n(z)=\frac{z^n}{i\sqrt{\pi}\Gamma(n+\frac{1}{2})2^n}\left\{F_n(z)+F_n(-z)\right\},
\ \ \ 
Y_n(z)  =  \frac{1}{i}\frac{z^n}{i\sqrt{\pi}\Gamma(n+\frac{1}{2})2^n}\left\{F_n(z)-F_n(-z)\right\}.
\eea
\\

A combination of Bessel functions that appears in the main text,
\begin{align}
&\frac{\frac{{\rm d}}{{\rm d}z}\left\{J_0(z)Y_0(za)-Y_0(z)J_0(za)\right\}}{J_0(z)Y_0(za)-Y_0(z)J_0(za)}
\nn\\
&=-\frac{a\,J_0(z)Y_1(za)-a\,Y_0(z)J_1(za)+J_1(z)Y_0(za)-Y_1(z)J_0(za)}{J_0(z)Y_0(za)-Y_0(z)J_0(za)},
\label{besselappendix}
\end{align}
 is re-expressed in terms of $F_n(z)$ as
 \begin{align}
(\ref{besselappendix})&=-\frac{a\,F_0(z)F_1(-za)-a\,F_0(-z)F_1(za)+F_1(z)F_0(-za)-F_1(-z)F_0(za)}
{F_0(z)F_0(-za)-F_0(-z)F_0(za)}.
\label{besselappendix2}
\end{align}
Comparing Eq.~(\ref{besselappendix2}) with the definition of $F_n$~(\ref{fdef}) and the expressions for the Hankel functions~(\ref{hankelappendix}), we find that for $0<a<1$,
\begin{align}
(\ref{besselappendix})&=-\frac{H_1^{(2)}(z)}{H_0^{(2)}(z)}-a\frac{H_1^{(1)}(za)}{H_0^{(1)}(za)}+O\left(e^{-2(1-a){\rm Im}(z)}\right) \ \ \ \ \ 
{\rm for} \ {\rm Im}(z)\to\infty,
\\
(\ref{besselappendix})&=-\frac{H_1^{(1)}(z)}{H_0^{(1)}(z)}-a\frac{H_1^{(2)}(za)}{H_0^{(2)}(za)}+O\left(e^{2(1-a){\rm Im}(z)}\right) \ \ \ \ \ \
{\rm for} \ {\rm Im}(z)\to-\infty.
\end{align}
\\


\end{document}